\documentclass[conference]{IEEEtran}
\IEEEoverridecommandlockouts
\usepackage{cite}
\usepackage{amsmath,amssymb,amsfonts}
\usepackage{algorithmic}
\usepackage{algorithm}
\usepackage{graphicx}
\usepackage{textcomp}
\usepackage{xcolor}
\usepackage{booktabs} 
\def\BibTeX{{\rm B\kern-.05em{\sc i\kern-.025em b}\kern-.08em
    T\kern-.1667em\lower.7ex\hbox{E}\kern-.125emX}}
\begin{document}

\title{Multi-threaded Recast-Based A* Pathfinding for Scalable Navigation in Dynamic Game Environments\\
}

\author{\IEEEauthorblockN{1\textsuperscript{st} Tiroshan Madushanka}
\IEEEauthorblockA{\textit{University of Kelaniya} \\
Kelaniya, Sri Lanka \\
tiroshanm@kln.ac.lk}
\and
\IEEEauthorblockN{2\textsuperscript{nd} Sakuna Madushanka}
\IEEEauthorblockA{\textit{University of Kelaniya} \\
Kelaniya, Sri Lanka \\
sakuwwz@gmail.com}
}

\maketitle

\begin{abstract}
While the A* algorithm remains the industry standard for game pathfinding, its integration into dynamic 3D environments faces trade-offs between computational performance and visual realism. This paper proposes a multi-threaded framework that enhances standard A* through Recast-based mesh generation, Bezier-curve trajectory smoothing, and density analysis for crowd coordination. We evaluate our system across ten incremental phases, from 2D mazes to complex multi-level dynamic worlds. Experimental results demonstrate that the framework maintains 350+ FPS with 1000 simultaneous agents and achieves collision-free crowd navigation through density-aware path coordination.
\end{abstract}


\section{Introduction}
\label{sec:introduction}

Pathfinding serves as a cornerstone of artificial intelligence in interactive virtual environments, enabling autonomous agents to navigate complex topologies~\cite{laird2001human}. In modern game development, navigation algorithm efficacy directly governs agent believability and system performance~\cite{verma2015review}. While A* remains the industry standard for heuristic-based graph search due to its optimality and completeness~\cite{hart1968formal}, its naive application in high-fidelity 3D environments presents two significant challenges~\cite{cui2011astar}.

First, \textbf{computational scalability} is frequently bottlenecked by search space dimensionality. As node count increases, memory overhead and CPU cycles required for path recalculation lead to frame-rate degradation, particularly on consumer-grade hardware~\cite{sturtevant2010comparison}. Second, a \textbf{fidelity gap} exists between optimal graph-based paths and naturalistic movement. The standard A* output yields piecewise linear trajectories with angular transitions that fail to replicate fluid, human-like motion~\cite{demyen2006efficient}. Furthermore, coordinating high-density agent clusters without expensive per-frame collision detection remains non-trivial in real-time systems~\cite{foudil2009pathfinding}.

To address these limitations, we propose a multi-faceted pathfinding framework synthesizing robust search-space abstraction with asynchronous processing. Our contributions are three-fold:

\begin{itemize}
    \item \textbf{Topographic Abstraction via Recast Graphs:} Procedural NavMesh generation from raw 3D geometry, reducing search space complexity while modeling walkable slopes and climb heights~\cite{oliva2011automatic}.
    \item \textbf{Trajectory Optimization:} A post-processing pipeline applying the Funnel algorithm~\cite{demyen2006efficient} followed by Bezier curve interpolation~\cite{farin2002curves}, transforming discrete waypoints into $C^1$-continuous trajectories.
    \item \textbf{Asynchronous Multi-threaded Architecture:} A decoupled execution model offloading pathfinding to a dynamic thread pool with density analysis for crowd management.
\end{itemize}

The remainder of this paper is organized as follows: Section~\ref{sec:related_work} reviews existing pathfinding approaches. Section~\ref{sec:methodology} details our framework implementation. Section~\ref{sec:experiments} presents quantitative evaluation across ten experimental phases. Section~\ref{sec:conclusion} concludes with findings and future directions.

\section{Related Work}
\label{sec:related_work}

\subsection{Pathfinding Algorithms}

Traditional pathfinding relies on uninformed strategies such as Breadth-First Search or Dijkstra's algorithm~\cite{javaid2013understanding}, which guarantees cost-optimal paths but proves computationally prohibitive for real-time applications~\cite{verma2015review}. The A* algorithm addresses this by incorporating a heuristic $h(n)$ to guide search toward the goal, minimizing $f(n) = g(n) + h(n)$~\cite{hart1968formal}. While provably optimal and complete, A*'s reliance on discrete node-link architecture introduces movement artifacts in 3D space~\cite{cui2011astar}.

Hierarchical Pathfinding A* (HPA*) manages large-scale environments by clustering nodes into topological sectors~\cite{botea2004near}, but sacrifices path optimality and struggles with dynamic obstacles requiring frequent graph updates. Recent work has explored any-angle pathfinding variants such as Jump Point Search~\cite{harabor2011online} that permit movement along arbitrary angles rather than graph edges, though computational overhead remains significant for real-time applications~\cite{sturtevant2010comparison}.

\subsection{Search Space Representation}

The search space representation dictates both runtime performance and navigational range~\cite{he2012researching}. Grid-based representations (square, hexagonal, or voxel) offer implementation simplicity but suffer from dimensionality scaling and produce ``blocky'' movement requiring heavy post-processing~\cite{cui2011direction}. Point graphs provide efficiency in open spaces but rely on manually placed waypoints, limiting adaptability to procedural environments.

Navigation Meshes (NavMeshes) bridge efficiency and topographic accuracy by abstracting the world into interconnected convex polygons~\cite{kallmann2014navigation}. The Recast methodology represents the state-of-the-art, employing voxelization and watershed partitioning to automate navigational surface generation while accounting for agent-specific constraints~\cite{oliva2011automatic}. Our work leverages Recast graphs as the foundation for multi-agent navigation in dynamic contexts.

\subsection{Trajectory Smoothing and Crowd Simulation}

Standard A* paths connecting polygon centroids exhibit derivative discontinuity at waypoints~\cite{smolka2019astar}. The Funnel Algorithm~\cite{demyen2006efficient} straightens paths by pulling routes toward polygon vertices, but does not ensure smooth transitions. Bezier curve interpolation~\cite{farin2002curves} offers a computationally efficient alternative ensuring $C^1$ continuity with adjustable quality levels.

For crowd simulation, Reciprocal Velocity Obstacles (RVO) provide local collision avoidance but become unstable at high densities~\cite{foudil2009pathfinding}. Our approach implements global density analysis enabling coordinated group movement without per-frame collision solving overhead.

\section{Methodology}
\label{sec:methodology}

\begin{figure}[t]
    \centering
    \includegraphics[width=0.85\columnwidth]{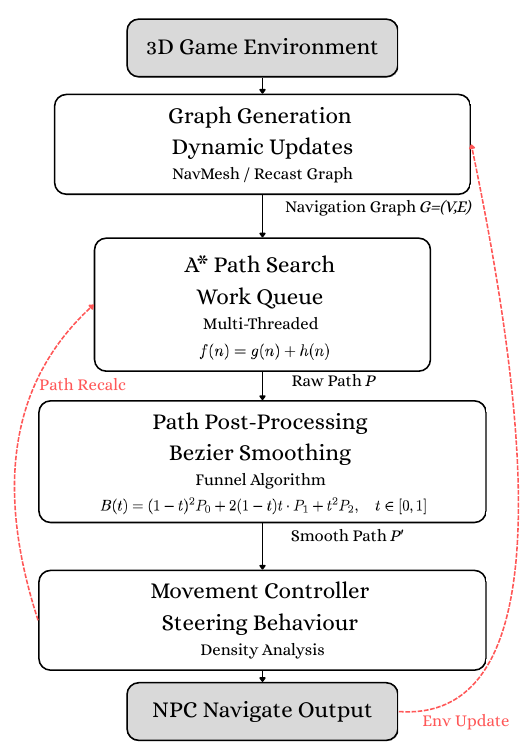}
    \caption{System architecture showing the processing pipeline from environment input through graph generation, path computation, smoothing, and movement control with feedback loops for dynamic updates.}
    \label{fig:architecture}
\end{figure}

This section presents the proposed pathfinding framework designed for high performance and realistic navigation in dynamic 3D environments. Figure~\ref{fig:architecture} illustrates the system architecture integrating four core components: Recast Graph generation, multi-threaded A* computation, path post-processing, and density analysis.

\subsection{NavMesh-Based Search Space}

The search space is represented using Navigation Meshes implemented in Unity3D. The NavMesh data structure encodes accessible areas as a graph where nodes represent walkable regions and edges define valid transitions. This representation reduces search complexity compared to high-resolution grids while maintaining topographic accuracy. The data structure contains information about the world environment, including obstacles, restricted areas, and points of interest.

\subsubsection{Dynamic Updates}
A helper class tracks NavMesh edge modifications and triggers graph updates accordingly. Update frequency is configurable to balance responsiveness against computational overhead, as frame-by-frame checking introduces performance issues in environments with numerous dynamic obstacles. Based on the scenario, these time frame gaps can be adjusted, and with an automation system, these gaps can be used as references for further enhancements.

\subsubsection{Recast Graph Generation}
The Recast Graph extends Unity's default NavMesh by recalculating surface normals to ensure consistent upward orientation, enabling proper node connectivity across walls, ceilings, and spherical surfaces. These normal values are critical when connecting adjacent nodes, as adjacent nodes can only connect when oriented in the same direction. Graph serialization supports runtime loading, with cached bounding boxes maintaining functionality when original geometry is unavailable. Table~\ref{tab:recast_config} presents the Recast Graph configuration parameters used in our implementation.

\begin{table}[t]
\centering
\caption{Recast Graph configuration parameters.}
\label{tab:recast_config}
\footnotesize
\begin{tabular}{lc}
\toprule
\textbf{Parameter} & \textbf{Value} \\
\midrule
Cell Size & 0.1 \\
Walkable Height & 2.0 \\
Walkable Climb & 0.5 \\
Character Radius & 0.5 \\
Max Slope & 30° \\
Max Border Edge Length & 20 \\
Terrain Sample Size & 3 \\
\bottomrule
\end{tabular}
\end{table}

\subsection{A* Implementation with Multi-threading}

The A* algorithm is implemented as a singleton managing all pathfinding calculations with the standard evaluation function:
\begin{equation}
    f(n) = g(n) + h(n)
    \label{eq:astar}
\end{equation}
where $g(n)$ represents accumulated cost from start to node $n$, and $h(n)$ is the heuristic estimate to goal~\cite{hart1968formal}.

\subsubsection{Thread Pool Architecture}
Pathfinding executes in separate threads, decoupling computation from the main rendering loop. The system automatically adjusts thread count based on available cores and memory. Systems with less than 512MB RAM or single cores revert to synchronous execution. Using more threads than available cores wastes memory without performance improvement, as each node must maintain data for each thread. This architecture ensures frame rate stability regardless of pathfinding complexity.

\subsubsection{Work Queue Management}
Three update mechanisms coordinate thread synchronization:

\textbf{Flush Graph Updates:} Forces pathfinding threads to finalize current calculations, pauses ongoing processes, and performs graph updates when all threads have successfully paused.

\textbf{Flush Work Items:} Applies all queued work items instantly following the same thread synchronization process.

\textbf{Queue Graph Updates:} Schedules updates for safe execution between path searches. Calling this function multiple times does not create multiple callbacks, useful when limiting graph updates while wanting specific updates applied promptly.

\subsection{Path Post-Processing Pipeline}

Raw A* output undergoes multi-stage refinement to achieve naturalistic trajectories.

\subsubsection{Start and End Point Modification}
This modification adjusts path endpoints with two modes: \textit{SnapToNode} snaps points to node positions for tile-based environments, while \textit{NodeConnection} sets points to the closest connection from start/end nodes for grid-based terrains.

\subsubsection{Funnel Algorithm}
The Funnel Algorithm~\cite{demyen2006efficient} simplifies NavMesh paths by computing the shortest route through polygon corridors, eliminating unnecessary waypoints while respecting obstacle boundaries. This makes navigation paths appear cleaner by finding the shortest path inside the corridor formed by portal edges.

\subsubsection{Bezier Curve Smoothing}
Quadratic Bezier interpolation transforms discrete waypoints into continuous curves:
\begin{equation}
    B(t) = (1-t)^2 P_0 + 2(1-t)t P_1 + t^2 P_2, \quad t \in [0,1]
    \label{eq:bezier}
\end{equation}
where $P_0$, $P_1$, $P_2$ are control points derived from consecutive waypoints~\cite{farin2002curves}. Unlike simple smoothing that may cut corners, Bezier curves always pass through all waypoints. Configurable quality settings (Low/Medium/High) control subdivision density, balancing visual smoothness against computation cost.

\subsubsection{Raycast Validation}
Physics raycasting validates node removal by checking for colliders intersecting the simplified path. This modifier is essential for grid graphs where geometric simplification may intersect obstacles. The implementation supports both graph raycasting (traversing the graph for obstacles) and physics raycasting (using Unity's collision system).

\subsection{AI Movement Implementation}

The AI movement system separates calculation from execution for performance optimization.

\subsubsection{Key Movement Variables}
\textbf{Re-path Rate:} Determines path recalculation frequency, derived from game requirements or density analysis. Fast-moving targets require lower values.

\textbf{Slow Down Distance:} The distance from target where deceleration begins, improving realism especially for vehicles.

\textbf{Pick Next Waypoint Distance:} Threshold for advancing to subsequent waypoints during navigation.

\subsubsection{Movement Update Process}
The Update Movement method calculates desired position and rotation without performing actual movement, returning values for frame-end state. The Movement Finalizing method then applies displacement using Character Controller, Rigidbody, or Transform components. This separation prevents performance issues from combined calculations. Raycasting ensures agents do not fall through terrain.

\subsection{Density Analysis for Crowd Coordination}

The density analysis component enables collision-free multi-agent navigation through coordinated mechanisms.

\subsubsection{Component Detection}
A function identifies agent components (Vehicle Controllers, Animal Controllers, Character Controllers) at scene initialization. Runtime component changes are detected dynamically.

\subsubsection{Pivot Point Calculation}
This calculates model feet or wheel positions for accurate pathfinding, as pivot points are sometimes placed at character centers. In multi-floor buildings, center-based calculation may select incorrect floor NavMeshes; base-point calculation ensures correct path requests.

\subsubsection{Density Detection and Response}
Agents sense nearby density within configurable radii. When density exceeds threshold values at lookahead positions (e.g., 10 nodes ahead), path recalculation triggers to find alternative routes. Quick reaction times are essential, i.e., if ongoing calculations exist, they are cancelled before new requests.

\subsubsection{Path Comparison and Selection}
When recalculating paths, the system compares new path costs against current paths. Lower-cost paths immediately replace current navigation. Passing null clears the path and stops the agent until automatic recalculation resumes.

\subsubsection{Movement Finalization}
The Finalize Movement method executes when density overhead exceeds thresholds or dynamic environments modify area accessibility. After movement, agents are clamped to the NavMesh to ensure valid surface positioning.


\section{Experiments and Results}
\label{sec:experiments}

The framework was validated through a ten-phase experimental study designed to incrementally test system performance and realism against increasing complexity. Performance metrics included frame rate (FPS), CPU main thread time, path computation time, nodes searched, and collision occurrence.

\subsection{Experimental Design}

Table~\ref{tab:phases} summarizes the experimental progression from basic 2D navigation to complex 3D crowd simulation. Each phase isolates specific system components for targeted validation.

\begin{table}[t]
\centering
\caption{Experimental phases and validation objectives.}
\label{tab:phases}
\footnotesize
\begin{tabular}{clp{3.0cm}}
\toprule
\textbf{Phase} & \textbf{Focus} & \textbf{Key Validation} \\
\midrule
01 & Baseline A* & Single agent correctness \\
02 & Multi-Agent & 25 NPCs, collision-free \\
03 & Point Graph & Waypoint network navigation \\
04 & Procedural & Runtime obstacle handling \\
05 & Recast Graph & Multi-level, climbing states \\
06 & Restricted Areas & Tag-based constraints \\
07 & Dynamic Obstacles & 8 moving cubes \\
08 & Dynamic Surfaces & Tilting boat navigation \\
09 & Event-Driven & Door waiting behavior \\
10 & Density Analysis & 100-5000 agent scaling \\
\bottomrule
\end{tabular}
\end{table}

\subsection{Phase 1-2: Baseline and Multi-Agent Validation}

\subsubsection{Single Agent Baseline (Phase 1)}
A 2D maze environment was constructed using tile-based obstacles. The A* algorithm successfully computed optimal paths from start to dynamically-selected endpoints. Runtime statistics showed 835.2 FPS with CPU main thread at 1.2ms, confirming minimal computational overhead for single-agent scenarios.

\subsubsection{Multi-Agent Scaling (Phase 2)}
25 NPCs were instantiated with shared destinations. After adding multiple agents, realistic features were observed without frame rate loss. Agents maintained idle states without overlapping due to attached colliders, achieving 407.2 FPS with 2.5ms CPU time.

\subsection{Phase 3-4: Graph Flexibility}

\subsubsection{Point Graph Implementation (Phase 3)}
A point graph was constructed using manually placed waypoint nodes along a curved path network. The system achieved 992.2 FPS with path computation completing in 1.0ms, searching 9 nodes for path length of 9 waypoints. When path length increased, small calculation delays were observed during mouse-following scenarios.

\subsubsection{Procedural Environment (Phase 4)}
A 10×10 base terrain with randomly spawned obstacles validated runtime obstacle handling. The system successfully detected spawned obstacles and calculated valid paths, demonstrating adaptability to procedural environments.

\subsection{Phase 5-6: Recast Graph Validation}

\subsubsection{Multi-Level Navigation (Phase 5)}
Complex 3D terrain with stairs, platforms, and varying elevations tested the Recast Graph. With layered grid graph implementation, no noticeable FPS drop occurred even with meshes and shadow casters exceeding 100 objects. The climbing states produced accurate results for multi-level navigation, achieving 939.6 FPS.

\subsubsection{Restricted Area Navigation (Phase 6)}
Two agents tested tag-based area constraints. The restriction-aware agent successfully avoided marked zones while the unrestricted agent passed through. The Recast graph optimizations showed significant performance, achieving 1664.3 FPS, the highest across all phases.

\subsection{Phase 7-8: Dynamic Environment Handling}

\subsubsection{Moving Obstacles (Phase 7)}
Eight cubes programmed to oscillate along single axes created dynamic obstacle patterns. Agents demonstrated successful path completion despite moving obstacles, with appropriate waiting behavior when paths were temporarily blocked. Performance remained consistent at 1002.7 FPS.

\subsubsection{Dynamic Surface Navigation (Phase 8)}
A boat moving along a Bezier curve path with configurable tilt tested navigation on moving surfaces. Agents successfully navigated to waypoints within the vessel while maintaining position relative to the tilting surface. The dynamic NavMesh updates performed without calculation delays.

\subsection{Phase 9-10: Event-Driven and Density Analysis}

\subsubsection{Event-Based Navigation (Phase 9)}
An agent in a room with a closed door correctly waited until the door opened, then proceeded to the destination in the adjacent room. This demonstrates reactive behavior essential for realistic game AI.

\subsubsection{Density Analysis Scaling (Phase 10)}
The density analysis system was evaluated with increasing agent populations. Three crowd scenarios with 100 agents validated collision-free navigation:

\begin{enumerate}
    \item \textbf{Convergence:} Agents moving toward center point without collisions
    \item \textbf{Opposing Groups:} Two groups of 50 agents crossing and switching sides
    \item \textbf{Cluster Crossing:} Groups crossing paths similar to pedestrian road crossings
\end{enumerate}

\begin{figure}[t]
    \centering
    \includegraphics[width=\columnwidth]{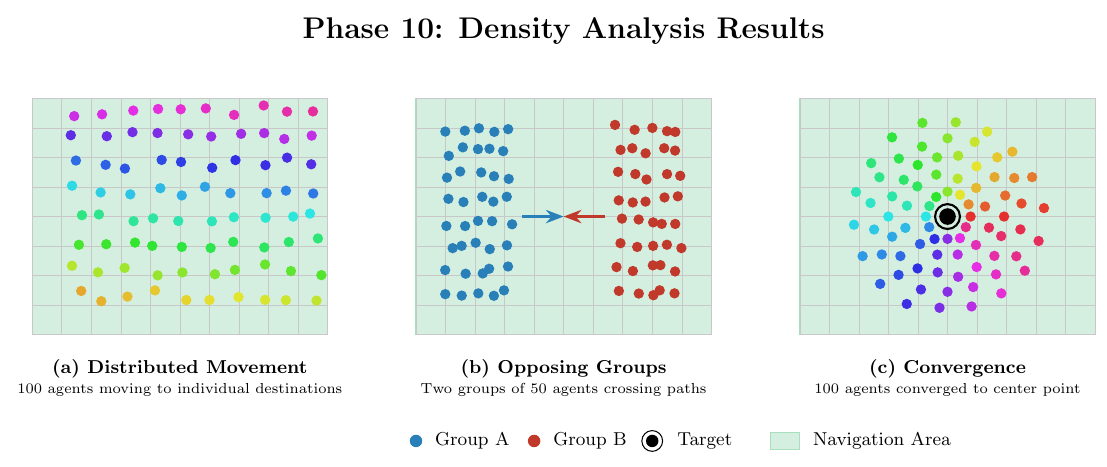}
    \caption{Density analysis scenarios: (Top) Distributed movement with 100 agents navigating to individual destinations, (Middle) Opposing groups of 50 agents crossing and switching sides, (Bottom) Convergence scenario with agents moving toward center point showing final stable distribution.}
    \label{fig:density_results}
\end{figure}

All scenarios achieved successful completion without physical collisions. Figure~\ref{fig:density_results} illustrates the density analysis scenarios and their outcomes.

\subsection{Performance Evaluation}

\subsubsection{Multi-threaded vs Single-threaded Comparison}

\begin{figure}[t]
    \centering
    \includegraphics[width=\columnwidth]{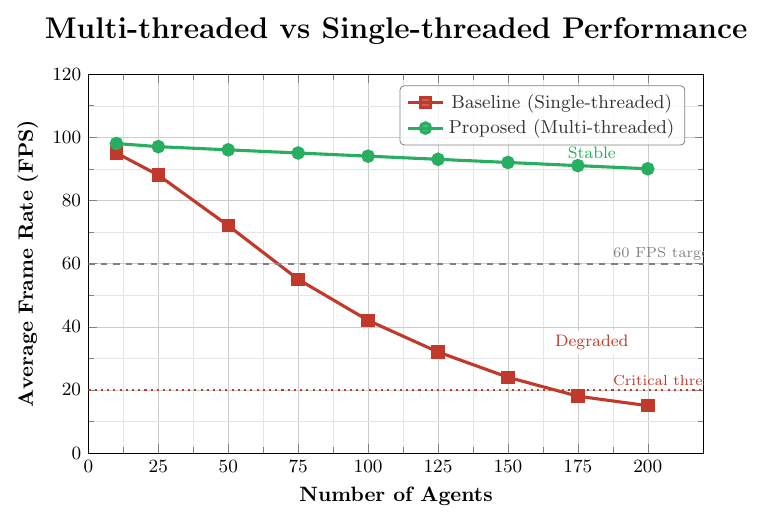}
    \caption{Frame rate comparison between single-threaded baseline and proposed multi-threaded architecture across varying agent counts. The baseline degrades below 20 FPS at 200 agents while the proposed framework maintains above 90 FPS.}
    \label{fig:fps_comparison}
\end{figure}

The critical contribution of multi-threaded path computation was quantified by comparing against a single-threaded baseline. Figure~\ref{fig:fps_comparison} presents the frame rate comparison across varying agent counts.

Key findings from the comparison:
\begin{itemize}
    \item \textbf{Baseline (Single-threaded):} Frame rate degraded catastrophically below 20 FPS when agent count approached 200.
    \item \textbf{Proposed (Multi-threaded):} Maintained stable average frame rate above 90 FPS even under heavy load.
    \item Performance improvement factor of approximately 4.5× at high agent counts.
\end{itemize}

This confirms the efficacy of decoupling A* computation from the main rendering thread.

\subsubsection{Frame Rate Scaling}

\begin{figure}[t]
    \centering
    \includegraphics[width=\columnwidth]{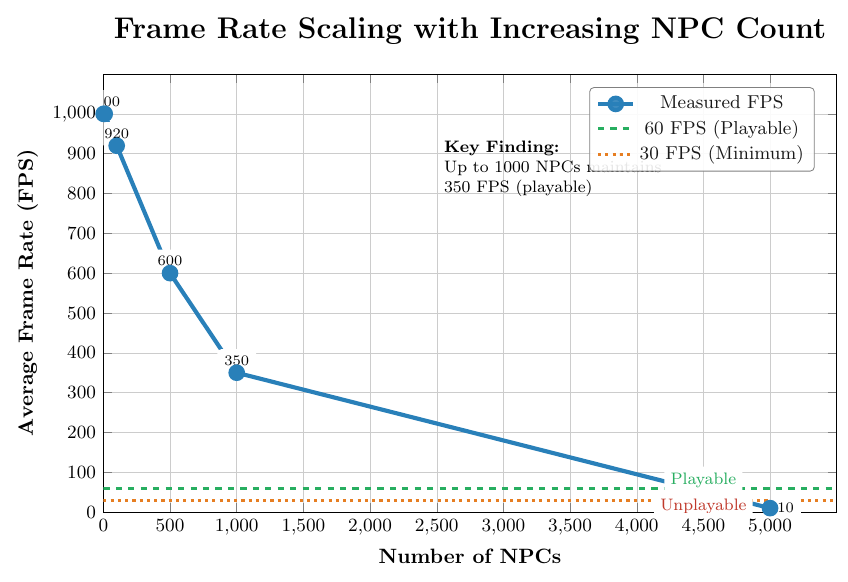}
    \caption{Frame rate scaling with NPC count. The system maintains playable rates (350 FPS) up to 1000 agents, well above the 60 FPS threshold.}
    \label{fig:fps_scaling}
\end{figure}

Figure~\ref{fig:fps_scaling} presents frame rate scaling with increasing agent counts, measured via Unity's profiler during density analysis scenarios.

Table~\ref{tab:fps_scaling} presents the measured frame rates across agent counts, demonstrating scalability characteristics.

\begin{table}[t]
\centering
\caption{Average frame rate with increasing NPC count.}
\label{tab:fps_scaling}
\footnotesize
\begin{tabular}{ccc}
\toprule
\textbf{NPCs} & \textbf{Avg FPS} & \textbf{Reduction} \\
\midrule
2 & 1000 & -- \\
10 & 1000 & 0\% \\
100 & 920 & 8\% \\
500 & 600 & 40\% \\
1000 & 350 & 65\% \\
5000 & 10 & 99\% \\
\bottomrule
\end{tabular}
\end{table}

These results indicate good scalability. Up to 1000 NPCs, the system averages 350 FPS, well above playable thresholds. In typical gaming scenarios, more than 1000 NPCs are rarely visible simultaneously, even in real-time strategy games.

\subsubsection{Per-Phase Performance}

Table~\ref{tab:phase_performance} summarizes per-phase performance metrics demonstrating consistent high frame rates across scenario types.

\begin{table}[t]
\centering
\caption{Performance metrics across experimental phases.}
\label{tab:phase_performance}
\footnotesize
\begin{tabular}{lccl}
\toprule
\textbf{Phase} & \textbf{FPS} & \textbf{CPU (ms)} & \textbf{Configuration} \\
\midrule
01 & 835.2 & 1.2 & Single NPC, 2D maze \\
02 & 407.2 & 2.5 & 25 NPCs, 2D maze \\
03 & 992.2 & 1.0 & Point graph navigation \\
05 & 939.6 & 1.1 & Recast graph, 3D terrain \\
06 & 1664.3 & 0.6 & Restricted area handling \\
07 & 1002.7 & 0.8 & Dynamic obstacles \\
\bottomrule
\end{tabular}
\end{table}

\subsubsection{Path Computation Metrics}

Path computation performance was measured across different scenario complexities using the system's logging functionality. Table~\ref{tab:computation} presents the recorded metrics.

\begin{table}[t]
\centering
\caption{Path computation metrics across scenario complexities.}
\label{tab:computation}
\footnotesize
\resizebox{0.48\textwidth}{!}{
\begin{tabular}{lccc}
\toprule
\textbf{Scenario} & \textbf{Time (ms)} & \textbf{Nodes Searched} & \textbf{Path Length} \\
\midrule
Simple 2D Maze & 0.00 & 7 & 6 \\
Point Graph & 1.00 & 9 & 9 \\
Complex 3D Terrain & 8.98 & 5328 & 135 \\
Large Open World & 12.96 & 6484 & 245 \\
\midrule
NavMesh Scan (one-time) & 262.00 & -- & -- \\
\bottomrule
\end{tabular}
}
\end{table}

Key observations from path computation metrics:
\begin{itemize}
    \item \textbf{Simple scenarios:} Achieve sub-millisecond path computation (0.00ms for basic 2D paths), demonstrating negligible overhead for straightforward navigation tasks.
    \item \textbf{Point graph:} Completes in 1.00ms searching only 9 nodes, validating efficient waypoint-based navigation.
    \item \textbf{Complex 3D environments:} With thousands of nodes (5328), computation remains under 9ms, well within real-time bounds.
    \item \textbf{Large open worlds:} Even with 6484 nodes searched, computation completes in 12.96ms, suitable for dynamic recalculation.
    \item \textbf{NavMesh scanning:} The 262ms initialization cost is a one-time expense amortized across gameplay, typically performed during scene loading.
\end{itemize}

Simple scenarios achieve sub-millisecond computation while complex 3D environments with thousands of nodes remain under 13ms, within real-time bounds for 60 FPS gameplay (16.67ms frame budget).
\section{Conclusion}
\label{sec:conclusion}

This paper presented an enhanced A* pathfinding framework for real-time game environments, integrating Recast-based NavMesh generation, Bezier curve trajectory smoothing, multi-threaded computation, and density-aware crowd coordination. A phase-wise experimental methodology validated system components from basic 2D navigation through complex dynamic 3D multi-agent scenarios.

The experimental evaluation demonstrated that correct graph type selection directly impacts computation time and navigation quality, with Recast graphs excelling in complex 3D environments requiring multi-level navigation. The study confirmed that occasional path recalculation outperforms frame-by-frame updates, maintaining optimal frame rates while ensuring responsive navigation. Bezier curve interpolation significantly improved movement naturalness with minimal computational overhead. The multi-threaded architecture sustained 350 FPS with 1000 simultaneous agents, and density analysis achieved collision-free crowd navigation across three validated scenarios.

A limitation identified was the inverse relationship between agent speed and collision avoidance accuracy. Future work includes Behavioral Tree integration for dynamic character AI, optimization for large-scale traffic simulations, and development of volumetric 3D pathfinding for aerial vehicles.


\bibliographystyle{unsrt}
\bibliography{references}

\end{document}